\documentclass[twocolumn,english,reprint, longbibliography, superscriptaddress, breaklinks=true, showkeys, showpacs=false, nofootinbib]{revtex4}
\usepackage[T1]{fontenc}
\usepackage[latin9]{inputenc}
\usepackage{color}
\usepackage{babel}
\usepackage{amsmath}
\usepackage{amssymb}
\usepackage{subfigure}
\usepackage{graphicx}
\usepackage{physics}
\usepackage{tabularray}
\usepackage{tabularx}
\usepackage{array}
\usepackage{float}
\usepackage{subcaption}

\usepackage[dvipsnames]{xcolor}

\DeclareMathAlphabet{\mymathbb}{U}{BOONDOX-ds}{m}{n}
\usepackage[unicode=true,pdfusetitle,
 bookmarks=true,bookmarksnumbered=false,bookmarksopen=false,breaklinks=true,pdfborder={0 0 0},backref=false,colorlinks=true]
 {hyperref} 
\makeatletter
\@ifundefined{textcolor}{}
{%
 \definecolor{BLACK}{gray}{0}
 \definecolor{WHITE}{gray}{1}
 \definecolor{RED}{rgb}{1,0,0}
 \definecolor{GREEN}{rgb}{0,1,0}
 \definecolor{BLUE}{rgb}{0,0,1}
 \definecolor{CYAN}{cmyk}{1,0,0,0}
 \definecolor{MAGENTA}{cmyk}{0,1,0,0}
 \definecolor{YELLOW}{cmyk}{0,0,1,0}
}

\pdfoutput=1
\hypersetup{colorlinks=true,citecolor=blue,linkcolor=cyan,urlcolor=blue,filecolor= green, breaklinks=true}
\usepackage{url}
\usepackage{breakurl}
\makeatother

\usepackage{mathrsfs}

\begin{document}

\author{Tiago Pernambuco\href{https://orcid.org/0009-0004-5692-0145}{\includegraphics[scale=0.05]{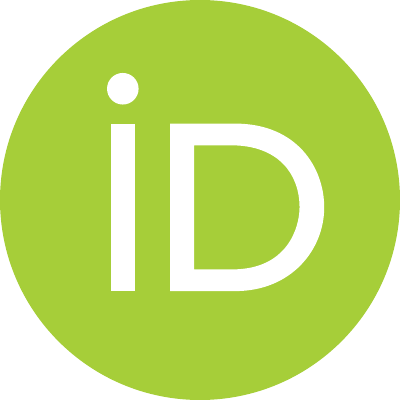}}}
\email{tiago4iece@gmail.com}
\affiliation{Theoretical and Experimental Physics Department, Federal University of Rio Grande do Norte, 59078-970, Natal, Brazil}

\author{Jonas Maziero\href{https://orcid.org/0000-0002-2872-986X}{\includegraphics[scale=0.05]{orcidid.pdf}}}
\email{jonas.maziero@ufsm.br}
\affiliation{Physics Department, 
Federal University of Santa Maria, 97105-900,
Santa Maria, RS, Brazil}

\author{Rafael Chaves\href{https://orcid.org/0000-0001-8493-4019}{\includegraphics[scale=0.05]{orcidid.pdf}}}
\email{rafael.chaves@ufrn.br}
\affiliation{International Institute of Physics, Federal University of Rio Grande do Norte, 59078-970, Natal, Brazil}
\affiliation{School of Science and Technology, Federal University of Rio Grande do Norte, Natal, Brazil}

\selectlanguage{english}

\title{Efficient detection of localization transitions using predictability}

\begin{abstract} 
Identifying phase transition points is a fundamental challenge in condensed matter physics, particularly for transitions driven by quantum interference effects, such as Anderson and many-body localization. Recent studies have demonstrated that quantum coherence provides an effective means of detecting localization transitions, offering a practical alternative to full quantum state tomography and related approaches. Building on this idea, we investigate localization transitions through complementarity relations that connect local predictability, local coherence, and entanglement in bipartite pure states. Our results show that predictability serves as a robust and efficient marker for localization transitions. Crucially, its experimental determination requires exponentially fewer measurements than coherence or entanglement, making it a powerful tool for probing quantum phase transitions. 
\end{abstract}


\date{\today}

\maketitle

\section{Introduction}
\label{intro}

The nature of light has been debated for centuries, with competing corpuscular and wave theories shaping the trajectory of scientific thought \cite{Baierlein}. The concept of wave-particle duality, eventually extended to matter, played a central role in the early development of quantum theory \cite{Eisberg, Wright}. In 1928, Niels Bohr proposed that a quantum system exhibits either wave-like or particle-like behavior depending on the experimental arrangement \cite{Bohr}. Later developments revealed that both behaviors can manifest simultaneously, though their respective strengths are constrained by the principle of complementarity -also known as duality relations- which formalize the trade-off between wave and particle characteristics \cite{Wootters, Greenberger, Jaeger, Englert}.

In parallel, advances in quantum information science have introduced powerful new frameworks for understanding a wide range of physical phenomena, from condensed matter systems \cite{Zeng} to black holes \cite{Harlow}. A key development in this domain is the formulation of mathematical resource theories \cite{Chitambar}, which have driven the systematic study of quantum resources across both theoretical and applied contexts \cite{Amico, DeChiara, Cao}. Among these, quantum coherence \cite{Streltsov} and quantum entanglement \cite{Horodecki} stand out as two of the most fundamental and extensively investigated resources.

Notably, while interferometric visibility has long served as the standard measure of wave-like behavior \cite{Born}, recent studies have shown that quantum coherence offers a more general and operationally meaningful quantifier \cite{Mishra, Diego2023}. In conjunction with this shift, the development of rigorous frameworks for quantifying particle-like and wave-like behavior \cite{Durr, Englert2008} - alongside the formulation of complementarity relations (CRs) and complete complementarity relations (CCRs) - has substantially deepened our understanding of wave-particle duality. Unlike traditional approaches that rely on the entire experimental setup, CRs and CCRs are defined for specific quantum state preparations \cite{Bera, Jakob, Angelo}, offering a more intrinsic perspective. Crucially, these relations have been shown to emerge directly from the foundational postulates of quantum mechanics, thereby addressing a long-standing question concerning the theoretical origin of the complementarity principle \cite{Basso2020a, Basso2020b, Basso2021, Basso2022}.

In this paper, we employ complementarity relations, specifically the notion of predictability that emerges from them, to analyze the localization transition that occurs in certain many-body systems, where mechanisms prevent thermalization, preserving memory of initial states and suppressing transport. Notable examples are Anderson localization (AL) \cite{Anderson, Evers, Segev, Soukoulis, Proof} and many-body localization (MBL) \cite{Zhao, Alet, Abanin, Nandkishore, Proof}, the first arising from disorder-induced elastic scattering in non-interacting systems, while MBL extends this behavior to interacting systems, with both phenomena observed in a variety of experimental platforms \cite{Billy, Cutler, Manai, Schreiber, Smith, Kohlert}. While both AL and MBL phases share several qualitative features \cite{Abanin, Proof}, the MBL phase can also exhibit distinct behavior due to the presence of interactions that allow for the transport of quantum correlations, leading to rich dynamical phenomena \cite{Abanin, Proof}. Identifying these distinguishing features is a central challenge for which tools from quantum information have proven effective. For instance, following a global quench, the bipartite entanglement entropy \cite{Two-site, Proof, Abanin} as well as mutual information \cite{De Tommasi} in the MBL phase tend to grow and spread over time, in contrast to the saturation observed in AL systems. Within this context, quantum coherence \cite{Chen} has emerged as a reliable and experimentally accessible quantity for distinguishing between the two phases. Here, we demonstrate that predictability -- the complementary dual of coherence and entanglement-- can also serve as a robust and efficient marker of localization in many-body systems. Crucially, its experimental estimation requires exponentially fewer measurements than coherence or entanglement, making it a powerful and scalable tool for probing quantum phase transitions.

The remainder of this article is organized as follows. In Sec. \ref{complementarity}, we discuss the basics of complementarity relations and introduce the measures of predictability, coherence, and entanglement that will be employed throughout the paper. In Sec. \ref{localization}, we introduce the Hamiltonian underlying the Anderson and many-body localization, highlighting their relevant differences. In Sec. \ref{sec:ES_qnets}, we show how the predictability provides a reliable benchmark to study and identify localization, also comparing it to coherence and entanglement. In Sec. \ref{diff_states}, we extend our results to different initial states, establishing the consistency of our method and giving insight into the quench dynamics of various systems. Finally, in Sec. \ref{sec:FR}, we discuss our findings and point out interesting directions for future research.

\section{Complementarity Relations}
\label{complementarity}

Quantum coherence, $C(\rho_A)$, allows us to formally quantify the waveness of a quantum state $\rho_A=Tr_B(|\Psi\rangle_{AB}\langle\Psi|)$ in relation to a reference observable $O=\sum_j o_j|\beta_j\rangle\langle\beta_j|$. For an observer who knows the reduced state $\rho_A$, the predictability function, $P(\rho_A)$, quantifies how well the observer can predict the result of a measurement of $O$. Furthermore, an observer who knows the joint state $|\Psi\rangle_{AB}$ can use the quantum entanglement $E(|\Psi\rangle_{AB})$ between $A$ and $B$ to predict the results of the measurements of $O$ by making measurements in the system $B$.  

The quantities mentioned in the last paragraph -- coherence, predictability, and entanglement -- cannot take arbitrary values for a given state preparation $|\Psi\rangle_{AB}$. For a bipartite pure state of two \( d \)-level systems, they are constrained by a complete complementarity relation (CCR) that takes the form  \cite{Basso2020b, Basso2021, Basso2022} 
\begin{equation}  
    C(\rho_A) + P(\rho_A) + E(|\Psi\rangle_{AB}) = \xi(d),
    \label{CCR}
\end{equation}  
where \( \xi(d) \) is a dimension-dependent constant.
By invoking the non-negativity of the entanglement, the corresponding complementarity relation (CR) is obtained:  
\begin{equation}  
    C(\rho_A) + P(\rho_A) \leq \xi(d).  
\end{equation}  
  
Complementarity relations as those above were obtained in the literature for several coherence functions and their corresponding particleness quantifiers. In this article, for simplicity, we consider quantifiers based on \( l_1 \)-norm coherence. In this framework, coherence, predictability, and entanglement are expressed in terms of the elements of the density matrix \( \rho^A_{jk} = \langle\beta_j|\rho_A|\beta_k\rangle \) as follows:  
\begin{align}  
    & C_{l_1}(\rho_A) = \sum_{j\ne k}|\rho^A_{jk}|, \\  
    & P_{l_1}(\rho_A) = d - 1 - \sum_{j\ne k}\sqrt{\rho^A_{jj} \rho^A_{kk}}, \\  
    & E_{l_1}(|\Psi\rangle_{AB}) = \sum_{j\ne k} \big(\sqrt{\rho^A_{jj} \rho^A_{kk}} - |\rho^A_{jk}|\big).  
    \label{l1}
\end{align}  

It is known from the literature that entanglement and coherence can be used to pinpoint localization transitions. In the sequence, we show that predictability can also be applied for that task, with the difference that this last function is much cheaper to estimate experimentally. 
For example, for $N$ two-level systems, the density matrix can be written as \cite{qtomo}:
\begin{equation}
\rho=\frac{1}{2^N}\sum_{j_1,j_2,\cdots,j_N=0}^3 S_{j_0 j_1 \cdots j_N}\sigma_1^{j_1}\otimes\sigma_2^{j_2}\otimes\cdots\otimes\sigma_N^{j_N},
\end{equation}
with $\sigma_s^{j_s}$ being a Pauli matrix: $\sigma_s^0=I_2,\ \sigma_s^1=\sigma_x,\ \sigma_s^2=\sigma_y,\ \sigma_s^3=\sigma_z$. So, in principle, to obtain the predictability experimentally, we need to measure $N$ observables $\sigma_1^{3}\otimes\sigma_2^{3}\otimes\cdots\otimes\sigma_N^{3}$. On the other hand, the experimental quantification of entanglement and coherence requires measurement of $\mathcal{O}(4^N)$ observables.

\section{Anderson and Many-Body Localization}
\label{localization}

The concept of Anderson localization (AL) was introduced by P. W. Anderson in his seminal 1958 paper \cite{Anderson}, and has since been a central topic in condensed matter physics \cite{Evers,Segev}. The Anderson model for a spinless electron on a $N$-site chain is described by the Hamiltonian \cite{Abanin}
\begin{equation}
H = J\sum_{i=1}^N \left(c^\dagger_{i+1}c_i + c^\dagger_ic_{i+1}\right) + W\sum_{i=1}^N \epsilon_i n_i,
    \label{AndersonModel}
\end{equation}
where $c_i^\dagger$, $c_i$ are fermion creation and annihilation operators acting on site $i$, $n_i = c^\dagger_i c_i$ is the local number operator on site $i$ and each $\epsilon_i$ is taken from a uniform distribution in $[-1, 1]$.

As the intensity of the disorder $W$ increases relative to the hopping amplitude $J$, the electron's wavefunction tends to localize around few lattice sites, leading to an absence of diffusion in the chain \cite{Anderson, Abanin}. A crucial fact to note is that in the one-dimensional Anderson model all states are exponentially localized \cite{Soukoulis}.

The lack of diffusion in Anderson-localized systems has a significant influence on the system's properties: since the electrons do not diffuse, they do not transport energy within the system \cite{Abanin}. As such, there cannot be any lasting electric currents or heat exchange in these so-called Anderson insulators. Since the efficient exchange of energy is necessary for thermalization within a system, Anderson-localized systems are unable to fully thermalize \cite{Abanin}.

It has been noted that, due to the lack of interactions and transport in Anderson insulators, quantum entanglement does not spread in these types of localized phases. Thus, after a global quench, a localization transition can be observed in non-interacting systems through a distinct saturation of entanglement entropies at (generally) finite values \cite{Zhao}.

\begin{figure}[H]
    \centering
    \includegraphics{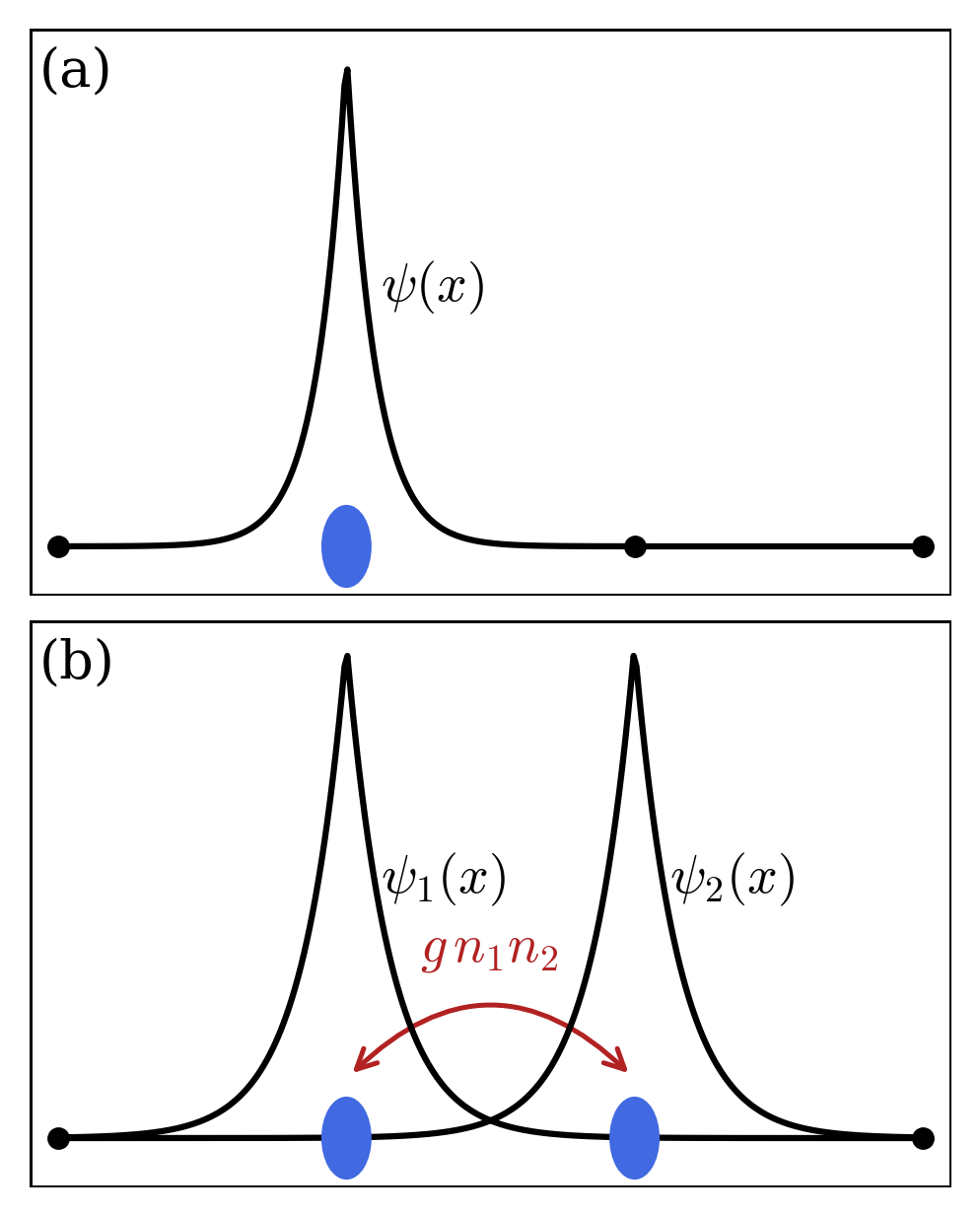}
    \caption{An illustration of (a) Anderson and (b) many-body localization, showing single-particle wavefunctions of fermions (blue circles). In (a) there is only a single, non-interacting fermion which is exponentially localized around the site $i = 1$. In (b) we have two fermions in neighboring sites ($i = 1$ and $j = 2$). Even if their single-particle wavefunctions are highly (although not as sharply as in the non-interacting case) localized, leading to lack of transport, the interactions between the particles cause entanglement to spread with time.}
    \label{AL_MBL}
\end{figure}

The Hamiltonian \eqref{AndersonModel} is quite idealistic, as it describes a non-interacting fermion model. As such, an obvious extension of the Anderson Model is to add a nearest-neighbor interaction term between the particles. This can be realized through a model of the following form \cite{Alet}:
\begin{equation}
H = J\sum_{i=1}^N \left(c^\dagger_{i+1}c_i + c^\dagger_ic_{i+1}\right) + g\sum_{i=1}^N n_i n_{i+1} + W\sum_{i=1}^N \epsilon_i n_i,
    \label{MBL}
\end{equation}
where $g$ is the strength of the interparticle interaction. Under sufficiently strong disorder, time evolution under \eqref{MBL} leads to many-body localized states, which have been found to hold several interesting properties such as the violation of the Eigenstate Thermalization Hypothesis and perpetual memory of initial conditions \cite{Nandkishore}.

The unusual properties described above have put many-body localized systems in the spotlight as candidates for the construction of quantum memories \cite{Smith, Nandkishore}, making a strong comprehension of many-body localization (MBL) fundamental for many potential applications in quantum information and related fields.

Although MBL phases share many similarities with Anderson insulators, for example, a lack of energy transport (and thus thermalization) and no conduction of electrical currents, there are still differences between them (see Fig. \ref{AL_MBL}). One key difference is that, since there are inter-particle interactions in MBL models, entanglement still spreads due to interaction-induced dephasing in the subsystems' reduced density matrices \cite{Abanin, Proof}.
\section{Localization identification using predictability}
\label{sec:ES_qnets}

Localization transitions in quantum systems have for years been probed through the behavior of quantum entanglement 
\cite{Basko, Gornyi, Pal, Nandkishore, Gullans, Proof}. In particular, it was shown in Ref. \cite{Proof} that, in the MBL phase, entanglement has a universal behavior of logarithmic change. It was also demonstrated in Ref. \cite{Two-site} that two-site entanglement provides a more efficient means of identifying MBL phases in experiments. More recently, Ref. \cite{Chen} has shown that the time evolution of quantum coherence after a quench under an AL or MBL Hamiltonian can be used to identify a localization transition in the system's dynamics.

Given the strong relationship between entanglement, coherence, and predictability given by the CCR in Eq. \eqref{CCR}, one can expect that predictability could also be a good probe for localization transitions. Furthermore, predictability $P$, being a function of the diagonal elements of the density matrix, is more easily measurable than both entanglement and coherence quantifiers, 
making it a more practical way to identify localization in experimental settings.

To show how predictability quantifiers can be used to identify localization, we follow up on the approach used in Ref. \cite{Chen}. Given an initial N\'eel state $|1010...10\rangle$, we compute the $l_1$-norm entanglement, coherence, and predictability 
throughout time evolution under the Hamiltonians \eqref{AndersonModel} and \eqref{MBL} and identify the localization transition as a shift from power-law growth to saturation (AL) or logarithmic growth (MBL).

In particular, we deal with two different situations: a trivial case in which we do not take a bipartition of the system, and thus entanglement does not play a role in the CCR \eqref{CCR}, and a bipartite case where it is important.

\subsection{Trivial Case}

\begin{figure*}[!ht]
    \centering
    \includegraphics[width=1.0\textwidth]{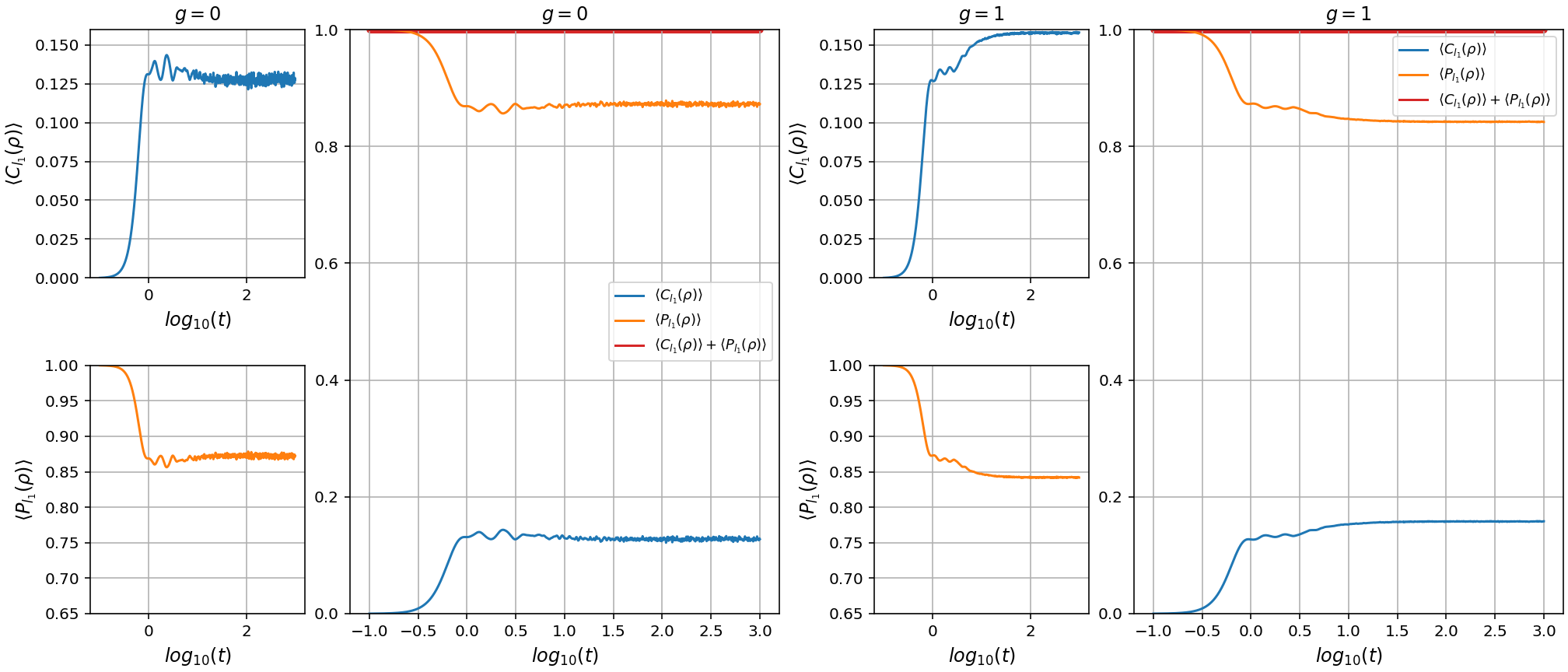}
    \caption{Disorder-averaged time evolution of Eqs. \eqref{CPTriv1}, \eqref{CPTriv2} and \eqref{SCR} for the Hamiltonian \eqref{MBL} with $J = 1, W = 2$. The values of $g$ (Anderson and MBL cases) are indicated above the subplots.}
    \label{W2}
\end{figure*}

Given a quantum state $\rho_{AB}$ as in the complementarity relation \eqref{CCR} take $A$ to be the entire system. Thus, there is no bipartition and $E(|\Psi_{AB}\rangle) = 0$. Denoting $\rho_{AB}$ by $\rho$ for simplicity, \eqref{l1} becomes
\begin{align}  
    & C_{l_1}(\rho) = \sum_{j\ne k}|\rho_{jk}|, \label{CPTriv1} \\  
    & P_{l_1}(\rho) = d - 1 - \sum_{j\ne k}\sqrt{\rho_{jj} \rho_{kk}},  
    \label{CPTriv2}
\end{align}  
and the complete complementarity relation reduces to a strict complementarity relation given by
\begin{equation}
C_{l_1}(\rho) + P_{l_1}(\rho) = \xi(d),
\label{SCR}
\end{equation}
where we choose a normalization such that $\xi(d) = 1$. This equality holds for a pure state $\rho$.

Given this setup and noting that $C_{l_1}$ was shown to identify the localization transition \cite{Chen}, $P_{l_1}(\rho) = 1 - C_{l_1}(\rho)$ trivially enables the same analysis. In any case, to verify and provide concrete evidence for this, we calculate the time evolution of an initial state $|1010...10\rangle$ under the Hamiltonians \eqref{AndersonModel} and \eqref{MBL} and calculate the coherences and predictability 
at each time step. More specifically, we always do so for one-dimensional systems of $N = 12$ sites and take disorder averages for our quantities of interest over $r = 100$ realizations. We consider three different set of parameters: $(J=1, W=2)$, $(J=1, W=6)$ and $(J=1, W=10)$ (always setting $g=1$ in the MBL model), with the results for the first case shown in  Fig. \ref{W2} and for the two latter cases in the Appendix \ref{sec:appendix}.

As shown in Fig. \ref{W2}, the analysis from Ref. \cite{Chen} can be directly extended to predictability. For the non-interacting model, the average predictability decreases from its initial value of 1 until reaching a minimum value and then stabilizes in the localized phase. For the interacting model, in contrast, despite following the same initial behavior, the average predictability eventually enters a logarithmic decrease regime, as opposed to the logarithmic growth of the average coherence \cite{Chen}. 

Thus, for the average global predictability, there are clear markers of localization in both interacting and non-interacting systems corresponding to stabilization and logarithmic change, as seen with coherence \cite{Chen} and entanglement \cite{Nandkishore, Basko, Gornyi, Pal, Gullans}.

\subsection{Bipartite Case}

Given a bipartite state $\rho_{AB}$, the CCR \eqref{CCR} holds. Since both $\langle C_{l_1}(\rho_{AB})\rangle$ \cite{Chen} and $\langle E_{l_1}(|\Psi_{AB}) \rangle$ \cite{Nandkishore, Basko, Gornyi, Pal, Gullans} present logarithmic behavior in the Anderson and many-body localized phases, one might expect that $P_{l_1}(\rho_{AB})$ should follow the same trend. To see why, let $C_{l_1}(\rho_{AB}) = A\log(Bt)$ and $E_{l_1}(|\Psi_{AB}) = C\log(Dt)$, where $A, B, C, D$ are arbitrary constants. Then, normalizing the complete complementarity relation so that $\xi(d) = 1$, we find that
\begin{align}
P_{l_1}(\rho_{AB}) = 1 - A\log(Bt) - C\log(Dt) 
\equiv a - b\log(t),
\end{align}
where $a$ and $b$ are constants.

\begin{figure*}[!ht]
    \centering
    \includegraphics[width=1.0\textwidth]{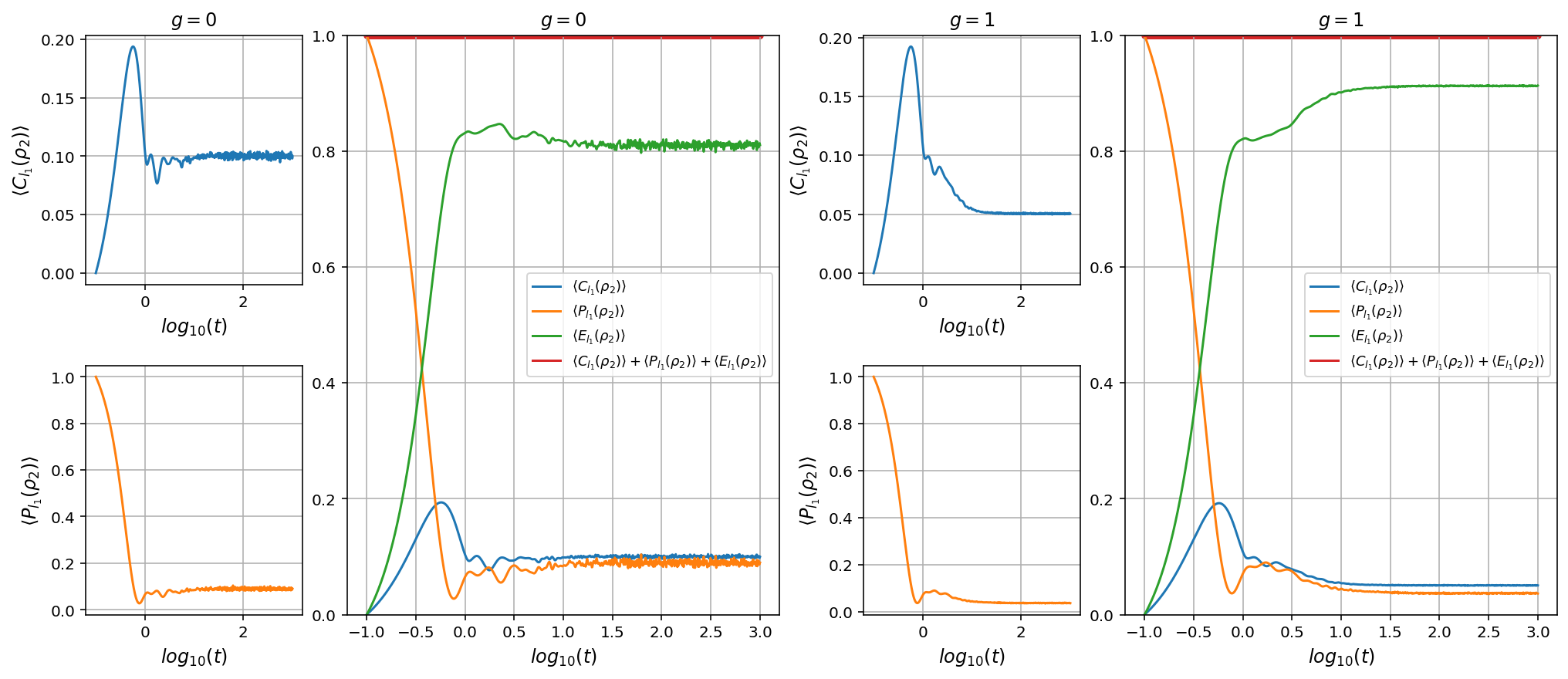}
    \caption{Time evolution of the average $2$-site local coherence, predictability, and entanglement entropy under the Hamiltonian \eqref{MBL} with $J = 1, W = 2$. The values of $g$ are, once again, displayed above the subplots.}
    \label{W2_bipartite}
\end{figure*}

To verify this behavior, we compute the $n$-site local coherence, defined as \cite{Chen}
\begin{equation}
    C_{l_1}(\rho_n) = \frac{1}{N - n +1}\sum_{i=1}^{N-n+1}C_{l_1}(\rho_{[i, i+1, ..., i+n-1]}),
    \label{localcoh}
\end{equation}
where we denote the reduced density matrix of the subsystem formed by sites $i, i+1,...,i+n-1$ by $\rho_{[i, i+1, ..., i+n-1]}$, and the corresponding local predictability given by
\begin{equation}
    P_{l_1}(\rho_n) = \frac{1}{N - n +1}\sum_{i=1}^{N-n+1}P_{l_1}(\rho_{[i, i+1, ..., i+n-1]}),
    \label{localpred}
\end{equation}
for a time evolution with the same sets of parameters as before and $n = 2$. We also compute the entanglement entropy $E_{l_1}$ from \eqref{CCR} and average all quantities over $r = 100$ realizations. As with the trivial case, the results for $W=2$ are shown in Fig. \ref{W2_bipartite}, while the results for $W=6$ and $W=10$ are shown in the Appendix \ref{sec:appendix}.

As can be seen from Fig. \ref{W2_bipartite}, we can once again extend the analysis from Ref. \cite{Chen} directly to the predictability: while in the Anderson-localized phase ($g = 0$) the average predictability stabilizes after some time, in the MBL phase it seems to display a very slow logarithmic change. In fact, it appears that, for the initial state we chose to compute the time evolution of, the changes in the coherence and entanglement entropy largely cancel out, leading to an almost constant predictability in the many-body localized phase.

\section{Results for Different Initial States}
\label{diff_states}

\begin{figure*}[!ht]
    \centering
    \includegraphics[width=1.0\textwidth]{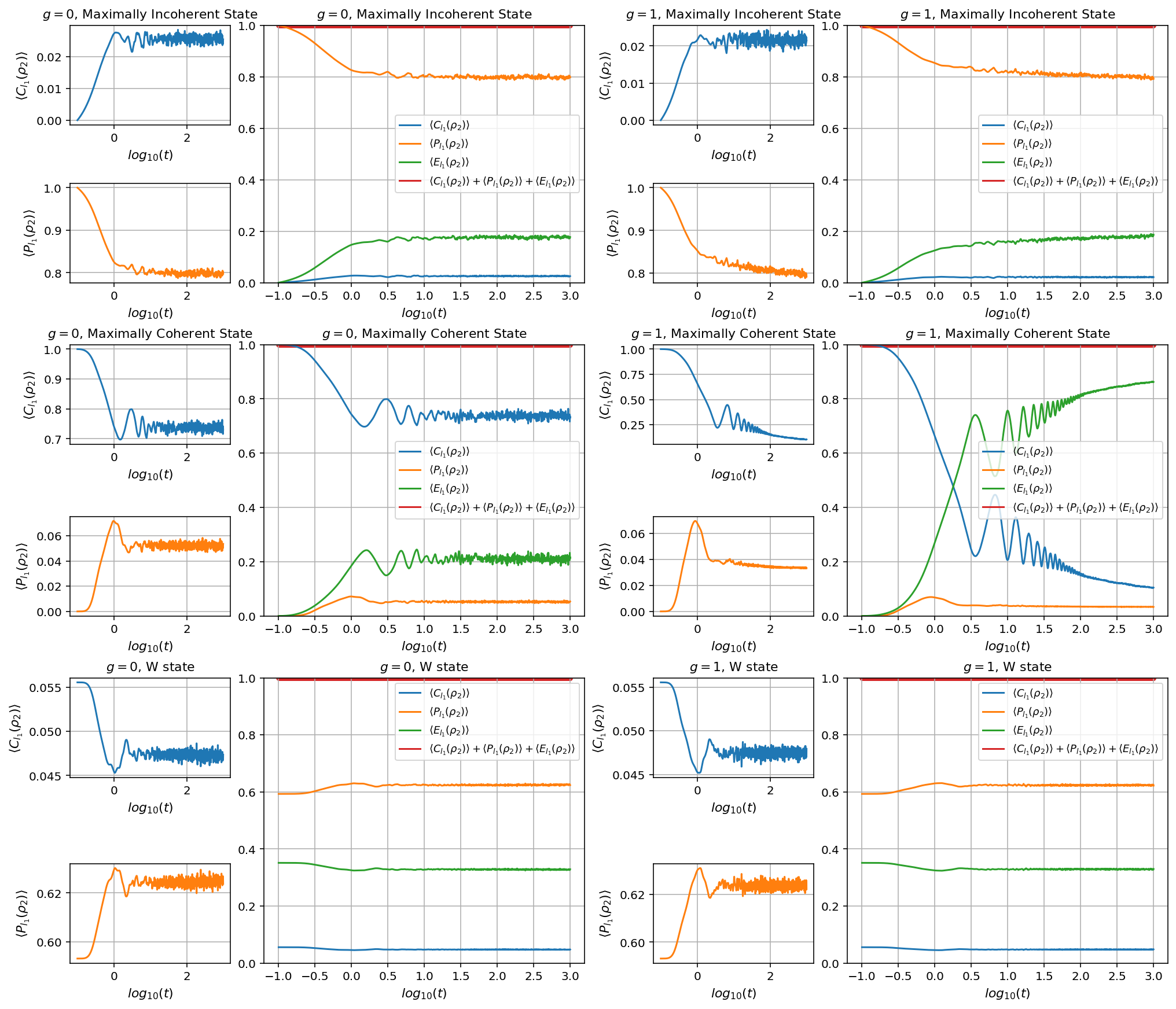}
    \caption{Time evolution of the average $2$-site local coherence, predictability, and entanglement entropy under the Hamiltonian \eqref{MBL} with $J = 1, W = 2$. The values of $g$ are, once again, displayed above the subplots.}
    \label{EstadosDiferentes}
\end{figure*}

In this section, we discuss the applicability of our results to other initial states. Namely, following the work of Ref. \cite{Chen}, we analyze the time-evolution behavior of a maximally incoherent fermion chain 
\begin{equation}
    |\Psi_{MI}\rangle = |111111000000\rangle,
    \label{MI}
\end{equation}
where all excited modes are in the left part of the lattice, and a maximally coherent state defined by
\begin{equation}
|\Psi_{MC}\rangle = \frac{1}{\sqrt{2^N}}\sum_{i=0}^{2^N-1}|i\rangle.
    \label{MaxCoh}
\end{equation}

Following that, we also analyze $N$-qubit W state, corresponding to a superposition of a single excitation in all modes:
\begin{equation}
|W\rangle = \frac{1}{\sqrt N}\left(|100...0\rangle + |010...0\rangle + ... + |000...1\rangle\right).
    \label{W}
\end{equation}

Unless otherwise stated, all results are for $W = 10$ and $N = 12$ in the non-trivial case, considering only local coherence and predictability, along with the entanglement. Our results for these states are shown in Fig. \ref{EstadosDiferentes}.

For the incoherent initial state \eqref{MI}, we see that, in the Anderson localized phase, the predictability saturates after a short time, while in the MBL phase it displays logarithmic behavior, as is to be expected. As such, the predictability here is again a good marker of localization.

As with the maximally incoherent fermion chain, the localization of the initial maximally coherent state \eqref{MaxCoh} is also captured by the predictability as it is by the coherence and by bipartite entanglement. We must note, however, that the state $|\Psi_{MC}\rangle$ is not a physically allowed fermionic state vector since it contains a superposition of states with even and odd fermion numbers, which is prohibited by fermion parity superselection rules \cite{SSR}.

The results for the $N$-qubit W state again seem to show strong differences between an initial transient behavior and the final saturated behavior. Since the W state is a superposition of single-particle states, it should come as no surprise that the value of $g$ bears no influence on the time evolution. Thus, we see that the coherence, predictability and entanglement all correctly identify the localized phase for this initial state.

\section{Discussion}
\label{sec:FR}

In this work, we have demonstrated the utility of predictability, a quantity emerging from complementarity relations, as a practical and efficient marker of localization in many-body quantum systems, offering significant advantages in terms of experimental feasibility. Unlike traditional approaches based on coherence and entanglement, which typically require an exponentially large number of measurements, predictability can be accessed with substantially fewer measurements, making it an attractive and scalable tool for probing localization phenomena.

Our analysis confirmed that predictability effectively captures key features of localization dynamics. In MBL systems, we observed that its temporal evolution mirrors known signatures from entanglement and coherence, such as the slow growth of correlations and the persistent memory of initial states. In contrast, Anderson-localized systems exhibited the expected saturation behavior, consistently reflected in predictability dynamics. Moreover, we showed that predictability remains a robust indicator across different initial state preparations, suggesting a state-independent reliability that makes it particularly promising for experimental scenarios where initial conditions may vary or be difficult to fully control.

Looking ahead, our results might open promising directions for future research. One is to explore whether predictability can serve as a useful probe in the study of quantum information scrambling \cite{Swingle, Alba, Iyoda, Hosur, Sekino}, where local information spreads rapidly across many degrees of freedom. Since scrambling is typically characterized by out-of-time-ordered correlators (OTOCs) \cite{Swingle, Alba, Iyoda, Hosur}, understanding whether predictability can capture aspects of this delocalization through a more experimentally accessible measure would be highly relevant. Another exciting possibility is to investigate predictability in the context of the eigenstate thermalization hypothesis (ETH) \cite{Deutsch}, which provides a foundation for understanding thermalization in isolated quantum systems. In this work, we have numerically demonstrated that predictability exhibits distinct behaviors in thermal and localized states. Given that predictability quantifies measurement bias or asymmetry, it would be valuable to investigate analytically whether its behavior in systems that satisfy ETH fundamentally differs from its behavior in MBL phases, where ETH is known to fail \cite{Alet, Abanin, Proof}.

\begin{acknowledgments}
This work was supported by the Simons Foundation (Grant Number 1023171, RC), the Financiadora de Estudos e Projetos (grant 1699/24 IIF-FINEP), and the National Council for Scientific and Technological Development (CNPq) under Grants No. 309862/2021-3, No. 409673/2022-6, No. 421792/2022-1, No. 307295/2020-6, and No. 403181/2024-0, and by the National Institute for the Science and Technology of Quantum Information (INCT-IQ) under Grant No. 465469/2014-0. We would also like to thank the High Performance Computing Center (NPAD/UFRN) for providing computational resources for the development of this work.
\end{acknowledgments}



\appendix

\section{Localization results for different disorder parameters}
\label{sec:appendix}
\begin{figure*}[t!]
    \centering
    \includegraphics[width=0.9\textwidth]{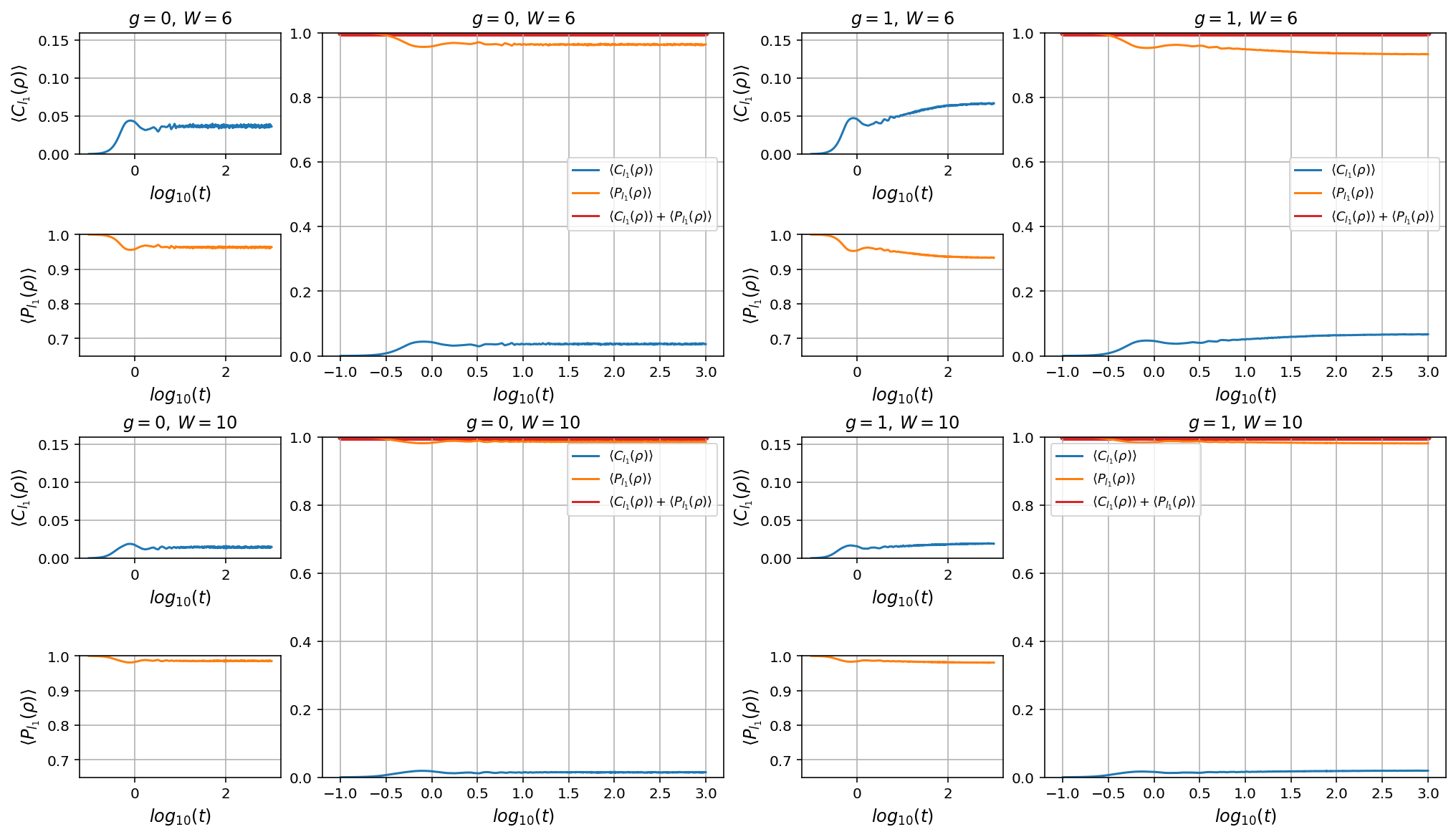}
    \caption{Disorder-averaged time evolution of Eqs. \eqref{CPTriv1}, \eqref{CPTriv2} and \eqref{SCR} for the Hamiltonian \eqref{MBL} with $J = 1$. The values of $W$ and $g$ (Anderson and MBL cases) are indicated above the subplots.}
    \label{AppTriv}
\end{figure*}

\begin{figure*}[t!]
    \centering
    \includegraphics[width=0.9\textwidth]{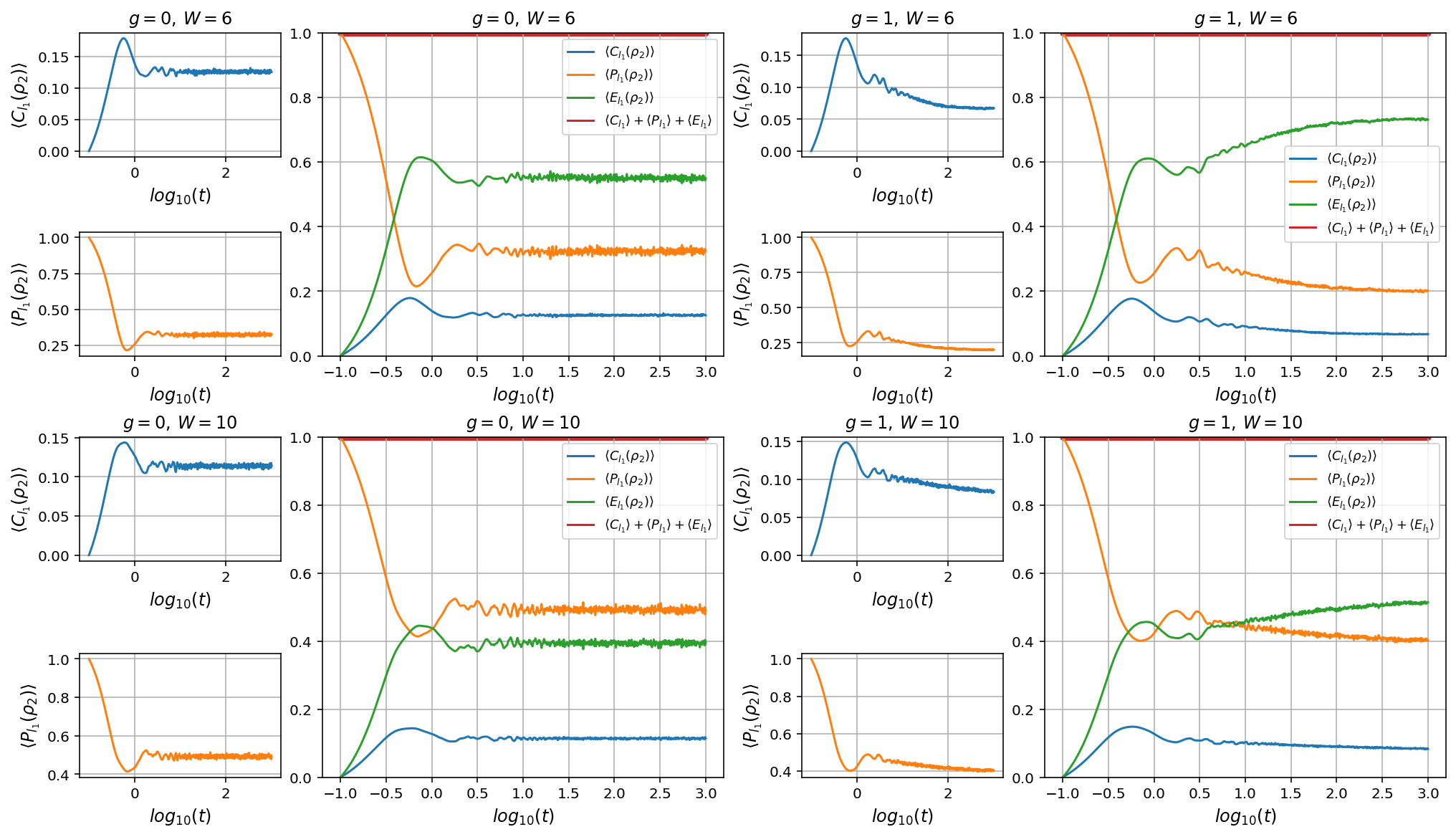}
    \caption{Time evolution of the average 2-site local coherence, predictability, and entanglement entropy under the Hamiltonian \eqref{MBL} with $J = 1$. The values of $W$ and $g$ are displayed above the subplots.}
    \label{AppBip}
\end{figure*}

To confirm that our findings are not sensitive to the specific choice of the disorder parameter $W$
in the Hamiltonians \eqref{AndersonModel} and \eqref{MBL}, we show that qualitatively similar results persist for different values of $W$. Fig. \ref{AppTriv} shows the result with partitioning the system (for which entanglement plays no role) and Fig. \ref{AppBip} refers to the bipartite case.

\end{document}